\documentclass[11pt]{article} 
\begin{document} 
  \vspace*{1.1cm} 
  \begin{center} 
  {\LARGE \bf The Paradox of Virtual Dipoles \\
  in the Einstein Action}
  \end{center} 

  \begin{center} 
  \vskip 10pt
  Giovanni Modanese$^*$
  \vskip 5pt
  {\it California Institute for Physics and Astrophysics \\
  366 Cambridge Ave., Palo Alto, CA 94306}
  \vskip 5pt
  and
  \vskip 5pt
  {\it University of Bolzano -- Industrial Engineering \\
  Via Sorrento 20, 39100 Bolzano, Italy}
  \end{center} 

  \baselineskip=.175in 
    
\begin{abstract}

The functional integral of pure Einstein 4D quantum gravity
admits abnormally large and long-lasting ``dipolar
fluctuations", generated by virtual sources with the
property $\int d^4x \sqrt{g(x)} {\rm Tr}T(x) = 0$.
These fluctuations would exist also at macroscopic scales,
with paradoxical consequences. We set out their general
features and give numerical estimates of possible
suppression processes.

\medskip
\noindent 04.20.-q Classical general relativity.

\noindent 04.60.-m Quantum gravity.

\end{abstract}

There has been considerable interest, in the last years, 
for the subject of vacuum fluctuations in 
quantum gravity. Several authors studied the possible
occurrence of large fluctuations in 2+1 dimensions,
in different contexts \cite{ash}. Other authors speculated
whether the Planck scale fluctuations typical of quantum
gravity, the so-called ``spacetime foam", generate a noise
which could be observed in certain circumstances \cite{ame}.
General phenomenological models for the effects of the
spacetime foam on high-energy scattering and dispersion
relations were also proposed \cite{ell}, based on the idea
that the amplitudes of these effects might be suppressed 
just by a $M_P^{-1}$ factor and not by $M_P^{-2}$.

In this work we describe a set of 
gravitational field configurations, called ``dipolar zero 
modes", which were not considered earlier in the 
literature. They give an exactly null contribution to the
Einstein action, being thus candidates to become large
fluctuations in the quantized theory. We give an explicit 
expression, to leading order in $G$, for some of the field 
configurations of this (actually quite large) set.
We also give an estimate 
of possible suppression effects following the addition 
to the pure Einstein action of cosmological or $R^2$
terms. This letter is based upon the paper \cite{vdf}, 
which in turn settles and refines some partial previous work 
\cite{pre}. 

These zero modes have two peculiar features, which make
them relatively easy to compute: (i) they are solutions of the
Einstein equations, though with unphysical sources;
(ii) their typical length scale is such that they can
be treated in the weak field approximation. We shall see
that these fluctuations can be large even on a
``macroscopic" scale. There are some, for instance, which
last $\sim 1 \ s$ or more and correspond to the field
generated by a virtual source with size $\sim 1 \ cm$
and mass $\sim 10^6 \ g$. This seems paradoxical, for
several reasons, both theoretical and phenomenological.
We have
therefore been looking for possible suppression
mechanisms. Our conclusion is that a vacuum energy term
$(\Lambda/8\pi G)\int d^4x \sqrt{g(x)}$ in the action
could do the job, provided it was scale-dependent and
larger, at laboratory scale, than its observed
cosmological value. This is at present only a speculative
hypothesis, however.

The dipolar fluctuations owe their existence to the fact
that the pure Einstein lagrangian $(1/8\pi G) 
\sqrt{g(x)} R(x)$ has indefinite sign also for static
fields. It is well known that the non-positivity of
the Einstein action makes an Euclidean formulation of
quantum gravity difficult; in that context, however,
the ``dangerous" field configurations have small
scale variations and could be eliminated, for instance,
by some UV cut-off. This is not the case of the dipolar
zero modes. They exist at any scale and do not make the
Euclidean action unbounded from below, but have instead
null (or $\ll \hbar$) action.

We shall consider the functional integral of pure quantum 
gravity, which represents a sum over all possible field 
configurations weighed with the factor $\exp[i\hbar
S_{Einstein}]$ and possibly with a factor due to the 
integration measure. The Minkowski space is a stationary 
point of the vacuum action and has maximum probability.
``Off-shell" configurations, which are not solutions of 
the vacuum Einstein equations, are admitted in the 
functional integration but are strongly suppressed by 
the oscillations of the exponential factor.

Due to the presence of the dimensional constant $G$ 
in the Einstein action, the most probable quantum 
fluctuations of the gravitational field ``grow" at very 
short distances, of the order of $L_{Planck}= \sqrt{G
\hbar/c^3} \sim 10^{-33} \ cm$. This led Wheeler, Hawking, 
Coleman and others to depict spacetime at the Planck 
scale as a ``quantum foam" \cite{haw}, with high curvature 
and variable topology. For a simple estimate
(disregarding of course the possibility of topology 
changes, virtual black holes nucleation etc.), suppose 
we start with a 
flat configuration, and then a curvature fluctuation 
appears in a region of size $d$. How much can the 
fluctuation grow before it is suppressed by the
oscillating factor $\exp[iS]$? (We set $\hbar=1$ and
$c=1$ in the following.) The contribution of the 
fluctuation to the action is of order $Rd^4/G$; both for 
positive and for negative $R$, the fluctuation is 
suppressed when this contribution exceeds $\sim 1$ in
absolute value, therefore $|R|$ cannot exceed $\sim 
G/d^4$. This means that the fluctuations of $R$ are 
stronger at short distances -- down to $L_{Planck}$, the 
minimum physical distance.
 
There is another way, however, to obtain vacuum field
configurations with action smaller than 1 in natural
units. Consider the Einstein equations 
and their covariant trace         
	\begin{eqnarray} 
	& & R_{\mu \nu}(x)- \frac{1}{2} g_{\mu \nu}(x) R(x)  
	= -8 \pi G T_{\mu \nu}(x); \label{ein} \\
	& & R(x)=8 \pi G {\rm Tr} T(x) =
	8 \pi G g^{\mu \nu}(x) T_{\mu \nu}(x).
\label{tra} 
\end{eqnarray} 
	Then consider a solution $g_{\mu \nu}(x)$ of equation 
(\ref{ein}) with a source $T_{\mu \nu}(x)$ obeying the 
additional integral condition 
	\begin{equation} 
	\int d^4x  \sqrt{g(x)}  {\rm Tr}  T(x) = 0.   
\label{add} 
\end{equation} 
	Taking into account eq.\ (\ref{tra}) we see 
that the Einstein action computed for this solution 
is zero. Condition (\ref{add}) can be satisfied by 
energy-momentum tensors that are not identically zero, 
provided they have a balance of negative and positive 
signs, such that their total integral is zero. Of course, 
they do not represent any acceptable physical source, but 
the corresponding solutions of (\ref{ein}) exist 
nonetheless, and are zero modes of the action.  
We shall give two explicit examples of virtual sources: 
(i) a ``mass dipole"
consisting of two separated mass distributions with
different signs; (ii) two concentric ``+/-
shells". In both cases there are some parameters of the
source which can be varied: the total positive and
negative masses $m_\pm$, their distance, the spatial
extension of the sources. 

Suppose we have a suitable source, with some free
parameters, and we want to adjust them in such a way
to generate a zero-mode $g_{\mu \nu}(x)$ for which
$S_{Einstein}[g]=0$. We shall always consider static
sources where only the component $T_{00}$ is non
vanishing. The action of their field is
	\begin{displaymath} 
	S_{zero-mode} = - \int d^4x \sqrt{g(x)}
	g^{00}(x) T_{00}(x).
\end{displaymath}

To first order in $G$, one easily finds in Feynman gauge
	\begin{displaymath}
	h_{\mu \nu}({\bf x})  
	= 2G (2\eta_{\mu 0} 
	\eta_{\nu 0} - \eta_{\mu \nu} \eta_{00})
	\int d^3y 
	\frac{T^{00}({\bf y})}{|{\bf x}-{\bf y}|}.
\end{displaymath}

It is straightforward to check that
$\sqrt{g(x)} g^{00}(x) = 1+o(G^2)$ and thus
$S_{zero-mode} = - \int d^4x T_{00}(x) + o(G^2)$.
Therefore provided the integral of the mass-energy density
vanishes, the action of our field configuration is of
order $G^2$, i.e., practically negligible, as we shall 
see now with a numerical example. Let us choose 
the typical parameters of the source as follows:
	\begin{equation}
	r \sim 1 \ cm; \qquad 
	m_\pm \sim 10^k \ g \simeq 10^{37+k} \ cm^{-1}
\label{esempio}
\end{equation}
	(implying $r_{Schw.}/r \sim 10^{-29+k}$). 
We assume in general an adiabatic switch-on/off of the
source, thus the time integral contributes to the action
a factor $\tau$. We shall keep $\tau$ (in natural units)
very large, in order to preserve the static character 
of the field. Here, for instance, let us take
$\tau \sim 1\ s \simeq 3 \cdot 10^{10} \ cm$.
With these parameters one finds
	\begin{displaymath}
	S_{zero-mode}^{order \ G^2} 
	\sim \tau \frac{G^2 m_\pm^2}{r^3}
	\sim 10^{-20+3k}.
\end{displaymath}  
	Thus the field generated by a virtual source 
with typical size (\ref{esempio}), satisfying the condition
$\int d^3x T_{00}({\bf x}) = 0$, has negligible action
even with $k=6$ (corresponding to apparent matter 
fluctuations with a density of $10^6 \ g/cm^3$ !) 
This should be compared to the huge action of the field
of a {\it single}, unbalanced virtual mass $m$; with 
the same values we have
$ S_{single \ m} = - \int d^4x \sqrt{g(x)} 
{\rm Tr} T(x) \sim \tau m + o(G^2) \sim 10^{47+k}$.

This numerical estimate shows that the cancellation of the first 
order term in $S_{zero-mode}$ allows to obtain a simple
lower bound on the strength of the fluctuations.
In principle, however, one could always find all the terms
in the classical weak field expansion, proportional to
$G$, $G^2$, $G^3$, etc., and adjust $T_{00}$ as to have 
$S_{zero-mode} =0$ exactly. They can be represented by
those Feynman diagrams of perturbative quantum gravity
which contain vertices with 3, 4 ... gravitons but do not 
contain any loops. 
The ratio between each contribution to $S$ and that of
lower order in $G$ has typical magnitude
$r_{Schw.}/r$,
where $r_{Schw.}=2\pi G m_\pm$ is the Schwarzschild radius
corresponding to one of the two masses and $r$       
is the typical size of the source. For a wide range of
parameters, this ratio is very small, so the expansion
converges quickly. 

As a first example of unphysical source satisfying 
(\ref{add}), consider the static field produced 
by a mass dipole. This consists of a positive source 
with mass 
$m_{+}$ and radius $r_{+}$ and a negative source with 
mass $-m_{-}$ and radius $r_{-}$, placed a distance 
$2a$ apart. The radii of the 
two sources are such that $a \gg r_\pm \gg r_{Schw.}$, 
where $r_{Schw.}$ is the Schwarzschild radius corresponding 
to the mass $m_{+}$. 
The action is found to be
$S_{Dipole} = - \int d^4x T_{00}({\bf x}) =
-\tau (m_+ - m_-) + o(G^2)$.
This vanishes for $m_+ = m_-$,
apart from terms of order $G^2$ (i.e., our dipoles have 
in reality a tiny monopolar component). The values of 
the masses and the radii $r_\pm$ (both of order $r$) can 
vary in a continuous way -- provided the condition above
is satisfied. Therefore these (non singular) ``dipolar" 
fields constitute a subset with nonzero volume 
in the functional integration. In fact, they are only 
a small subset of all solutions of the Einstein equations
with sources satisfying eq.\ (\ref{add}).

Another example is given by 
two concentric spherical shells, the internal
one with radii $r_1$, $r_2$, and the external one with
radii $r_2$, $r_3$ ($r_1 < r_2 < r_3$). Let the internal
shell have mass density $\rho_1$ and the external shell
density $\rho_2$, with opposite sign. 
The condition for zero action requires,
up to terms of order $G^2$, that the total positive mass
equals the total negative mass, i.e.,
$\rho_1 (r_2^3 - r_1^3) + \rho_2 (r_3^3 - r_2^3) = 0$.
The spherical symmetry of this source
offers some advantages in the calculations.

One may think that large gravitational fluctuations, if real, 
would not remain unnoticed. Even though vacuum fluctuations
are homogeneous, isotropic and Lorentz-invariant, they
could manifest themselves as noise of some kind. 
Most authors are skeptic about the possibility
of detecting the noise due to spacetime foam \cite{ame,ell},
but the virtual dipole fluctuations described in this
paper are much closer to the laboratory scale.
Observable quantities, like for instance the 
connection coefficients $\Gamma^\rho_{\mu \nu}$
could then exhibit strong fluctuations.

The existence of these
fluctuations would be paradoxical, however, already 
at the purely conceptual level. Common wisdom in particle
physics states that the vacuum fluctuations in free space
correspond to virtual particles or intermediate states 
which live very short, i.e., whose
lifetime is close to the {\it minimum} allowed by the Heisenberg
indetermination relation. 

Let us estimate the product
$E\tau$ for the dipolar fluctuations. The total energy of
a static gravitational field configuration vanishing at
infinity is the ADM energy. Since the source of a
dipolar fluctuation satisfies the condition $\int d^3x
T_{00}({\bf x})=0$ up to terms of order $G^2$, the
dominant contribution to the ADM energy is the Newtonian
binding energy \cite{mur}.

The binding energy of the 
field generated by a source of mass $m$ and size
$r$ is of the order of $E \sim -Gm^2/r$, where
the exact proportionality factor depends on the details
of the mass distribution. 
For a dipolar field configuration characterized by
masses $m_+$ and $m_-$ 
and radii of the sources $r_+$ and $r_-$,
the total gravitational energy is of the order of
$E_{tot} \sim - G m_\pm^2 ( r_-^{-1} + r_+^{-1})$
(disregarding the interaction energy between the two
sources, proportional to $1/a \ll 1/r$).
With the parameters (\ref{esempio}) we have 
$E_{tot} \sim -G m_\pm^2 \sim
-10^{12+k} \ cm^{-1}$. Remembering that 
$k$ can take values up to $k=6$,
we find for these dipolar fluctuations $\tau |E_{tot}| \sim
10^{28}$! (For comparison, remember the case of a
``monopole" fluctuation of virtual mass $m$ and duration $\tau$. The 
condition $S<1$ implies $\tau m<1$. The dominant contribution
to the ADM energy is just $m$, so the rule $E\tau<1$ is
respected.)

The Newtonian binding energy of the concentric +/- shells
turns out to be of the same magnitude order, more exactly
$E = \frac{Gm_\pm}{r} P(\beta)$,
where $P(\beta) \equiv P(r_3/r_2)$ is a polynomial which 
is positive
if $|\rho_1|>|\rho_2|$ (the repulsion between the two shells
predominates) and negative if $\rho_1<\rho_2$ (the 
attraction inside each shell predominates).
From the physical point of view it is reasonable
to admit -- remembering that we are in a weak field regime
and forgetting general covariance for a minute -- that
the binding energy is localized within the surface of
the outer shell (the field is $o(G^2)$ outside). The
energy density is therefore of the order of
$\frac{|E|}{r^3} \sim \frac{Gm_\pm}{r^4} \sim 10^{29+k} \ 
cm^{-4}$ (with the parameters (\ref{esempio})), and can take both
signs. This value looks quite large, even though the
Ford-Roman inequalities \cite{for} or similar bounds do
not apply to quantum gravity, where the metric is not 
fixed but free to fluctuate, and there is in general no 
way to define a local energy density.

Concerning possible suppression processes of the
dipolar fluctuations, here we just
quote the results. The contribution to the cosmological
term is $\Delta S_{\Lambda} \sim \tau \Lambda m_\pm r^2
\sim 10^{-3+k}$, with the parameters above, and the
contribution to an $\alpha R^2$ term is of the order
of $\tau G^2 m_\pm^2/r^3 \sim \alpha 10^{-48+2k}$. We see
that only the cosmological term can act as a cut-off at
macroscopic scales.

\medskip
{\bf Acknowledgment} - This work was supported in part by the California Institute for Physics
and Astrophysics via grant CIPA-MG7099.

\medskip
\noindent $^*$ e-mail address: giovanni.modanese@unibz.iz


\begin{thebibliography}{9}


\bibitem{ash}
A.\ Ashtekar, Phys.\ Rev.\ Lett.\ {\bf 77}, 4864 (1996).
R.\ Gambini and J.\ Pullin, Mod.\ Phys.\ Lett.\ {\bf A 12}, 
2407 (1997).
A.\ E.\ Dominguez and M.\ H.\ Tiglio, Phys.\ Rev.\ 
{\bf D 60}, 064001 (1999).

\bibitem{ame}
G.\ Amelino-Camelia, Nature {\bf 398}, 216 (1999); 
Phys.\ Lett.\ {\bf B 477}, 436 (2000).
R.\ J.\ Adler, I.\ M.\ Nemenman, J.\ M.\ Overduin and D.\ 
I.\ Santiago, Phys.\ Lett.\ {\bf B 477}, 424 (2000).

\bibitem{ell}
A.\ A.\ Kirillov, Sov.\ Phys.\ JETP (J.\ Exp.\ Theor.\
Phys.) {\bf 88}, 1051 (1999). 
J.\ Ellis, N.\ E.\ Mavromatos and D.\ V.\ Nanopoulos,
Phys.\ Rev.\ {\bf D 61}, 027503 (2000).

\bibitem{vdf}
G.\ Modanese, {\it Large ``Dipolar" Vacuum Fluctuations 
in Quantum Gravity} (gr-qc/0005009).

\bibitem{pre}
G.\ Modanese, Phys.\ Rev.\ {\bf D 59}, 024004 (1998);
Phys.\ Lett.\ {\bf B 460}, 276 (1999).

\bibitem{haw}
J.A.\ Wheeler, Ann.\ Phys.\ {\bf 2} (1957) 604.
S.W.\ Hawking, Nucl.\ Phys.\ {\bf B 144}, 349 (1978). 
S.\ Coleman, Nucl.\ Phys.\ {\bf B 310}, 643 (1988).

\bibitem{mur}
N.O.\ Murchadha and J.W.\ York, Phys.\ Rev.\ {\bf D 10} 
(1974) 2345.


\bibitem{for}
L.H.\ Ford and T.A.\ Roman, Phys.\ Rev.\ {\bf D 43}, 3972 
(1991); {\bf D 46}, 1328 (1992); {\bf D 51}, 4277 (1995); 
{\bf D 55}, 2082 (1997).


\end{thebibliography}
\end{document}